\documentclass[10pt, conference]{IEEEtran}
\IEEEoverridecommandlockouts
\usepackage{tablefootnote}

\usepackage[utf8]{inputenc}
\usepackage{textcomp}
\usepackage{url}
\usepackage{cite}
\usepackage{amsmath,amssymb,amsfonts}
\usepackage{algorithmic}
\usepackage{algorithm}
\usepackage{comment}
\usepackage{graphicx}
\usepackage[dvipsnames]{xcolor}
\usepackage{textcomp}
\usepackage{xcolor}
\usepackage{romannum}
\usepackage{multirow}
\usepackage{comment}
\def\BibTeX{{\rm B\kern-.05em{\sc i\kern-.025em b}\kern-.08em
    T\kern-.1667em\lower.7ex\hbox{E}\kern-.125emX}}
\usepackage[a4paper, total={184mm,239mm}]{geometry}
\def\BibTeX{{\rm B\kern-.05em{\sc i\kern-.025em b}\kern-.08em
    T\kern-.1667em\lower.7ex\hbox{E}\kern-.125emX}}

\IEEEpubid{\makebox[\columnwidth]{ 979-8-3315-2710-5/25/ \$31.00 ©2025 IEEE  \hfill{ \hspace{\columnsep} \makebox[\columnwidth]{ }}}} 
    
\begin{document}

\title{A Compact, Low Power Transprecision ALU for Smart Edge Devices\\

\thanks{This work was supported in part by the NSF I/UCRC IDEAS center and from the NSF grants \#2324945 and \#2425535.}
}

\author{\IEEEauthorblockN{Ayushi Dube}
\IEEEauthorblockA{
\textit{Arizona State University}\\
Tempe, AZ, USA \\
adube9@asu.edu}
\and
\IEEEauthorblockN{Gian Singh}
\IEEEauthorblockA{
\textit{Arizona State University}\\
Tempe, AZ, USA  \\
gsingh58@asu.edu}
\and

\IEEEauthorblockN{Sarma Vrudhula}
\IEEEauthorblockA{
\textit{Arizona State University}\\
Tempe, AZ, USA \\
svrudhul@asu.edu}
\and
}

\maketitle

\begin{abstract}
Transprecision computing (TC) is a promising approach for energy-efficient machine learning (ML) computation on resource-constrained platforms. This work presents a novel ASIC design of a Transprecision Arithmetic and Logic Unit (TALU) that can support multiple number formats: Posit, Floating Point (FP), and Integer (INT) data with variable bitwidth of 8, 16, and 32 bits. Additionally, TALU can be reconfigured in runtime to support TC without overprovisioning the hardware. Posit is a new number format, gaining traction for ML computations, producing similar accuracy in lower bitwidth than FP representation. This paper thus proposes a novel algorithm for decoding Posit for energy-efficient computation. TALU implementation achieves a 54.6$\times$ reduction in power consumption and 19.8$\times$ reduction in the area as compared to a state-of-the-art unified MAC unit (UMAC)~\cite{2022_Crespo_IEEETCS_UM} for Posit and FP computation. Experimental results on an ML compute kernel executed on a Vector Processor of TALUs integrated with a RISC-V processor achieves about 2$\times$ improvement in energy efficiency and similar throughput as compared to a state-of-the-art TC-based vector processor.

  
\end{abstract}


\section{Introduction}

Transprecision computing (TC) dynamically adjusts the number format and precision of computations to balance numerical accuracy and energy efficiency~\cite{2021_Mach_TVLSI}. This approach has proven highly effective in improving the energy efficiency of compute-intensive applications on resource-constrained platforms without compromising the accuracy~\cite{2021_Mach_TVLSI,2018_Malossi_DATE,2020_Grosser_ACM}. TC is extensively used in applications such as ML (INT4, INT8, INT16, FP16, FP32) and cryptography~\cite{2018_Chen_CryptographyBook} (INT32, INT64). Therefore, hardware support for TC is crucial for power-constrained edge devices.

Besides controlling the bitwidth, the choice of the number format such as the Floating Point (FP), Integer (INT) and Posit plays an important role in determining the accuracy and energy efficiency of the application as each number format supports a unique dynamic range of values. 

Posit is a new number format that addresses some inherent problems in FP~\cite{2023_Leong_Springer}. It results in higher accuracy, energy efficiency, and robustness than FP for many ML workloads \cite{2021_Romanov_IEEEAccess}. However, it cannot simply replace FP, as there are numerous applications where INT and FP are critical for general-purpose CPU computations and other specialised tasks. Hence, to realise the full benefits of TC, it is essential to support INT, FP, and Posit in a single compute unit or core.

TC can be used at different granularities depending on the application requirements. From the \textit{node-level}, where a node can be a scalar operation or a macro like matrix multiplication ~\cite{2022_Dube_ICCAD}, to the \textit{layer-level} where a layer is a neural network layer ~\cite{2023_Baleen_ArXiv,2024_SHo_DeepSeek}. Implementing TC at multiple levels of granularity on a single hardware platform is not simple due to the inherent trade-off between high energy efficiency and computational flexibility~\cite{2018_Malossi_DATE}. Therefore, developing efficient architectures that support \textit{multi-granularity} TC remains an essential goal to improve energy efficiency and performance of computing systems.

This work presents a novel design of a transprecision arithmetic and logic unit, named TALU, that supports multiple data formats, including  Posit, FP, and INT, for different precisions, with minimal hardware overhead. To achieve multi-granularity TC, TALU provides reconfigurability to support different data formats at the basic arithmetic and logic operation level. This reconfiguration control can be employed at the \textit{node level} or at the \textit{layer level} according to the application requirements.

An SIMD core, named TALU-V, consisting of a vector of TALUs, is designed and integrated with a lightweight RISC-V processor (RISCY \cite{2017_Gautschi_TVLSI}). A TALU-V can execute multiple scalar operations in parallel or a single vector operation and can serve as an efficient ML accelerator. TALU and TALU-V are specifically tailored for ultra-low power, ultra-compact smart devices operating in highly resource-constrained environments. The objective is to enable such devices to efficiently execute compute-intensive ML inference tasks at the edge.

The main contributions of this paper are summarized below.
\begin{itemize}

     \item A novel decoding algorithm is proposed for a Posit $P(n,e)$ that outputs the values of the fields in $P$, i.e., sign (S), regime (R), exponent (E), and mantissa (F), in a fixed number of cycles on TALU using parallel operations.
    
    \item TALU supports Posit, FP, and INT numbers with varying bitwidths (4 to 32-bits) in a substantially lower area and power than existing state-of-the-art designs. TALU is 20$\times$ smaller, consumes 54.6$\times$ lower power, and has 2.76$\times$ lower power density than state-of-the-art  FP and Posit MAC units~\cite{2022_Crespo_IEEETCS_UM}. TALU can compute multiple functions, unlike other existing designs that are only multiply and accumulate units (MAC) \cite{2022_Crespo_IEEETCS_UM,2019_Zhang_TC,2021_Zhang_ISCAS,2020_Neves_SiPS}.

    \item TALU compared with existing Posit-only compute elements for 32 bit computation exhibits 5.4$\times$ to 16.7$\times$ smaller area, 15.16$\times$ to 42.5$\times$ lower power and 2.53$\times$ to 4.13$\times$ lower power density. This indicates that TALU is ideal for low power edge applications. 
    
    \item A low power and low area processor architecture is proposed using a lightweight RISC-V microarchitecture, RISCY, tightly integrated with a vector unit of $N$ TALUs (TALU-V), named TALU-V. RISCY+TALU-V architecture executes small ML workloads at the edge with high energy efficiency. RISCY+TALU-V compared with a custom RISC-V based vector processor has 1.98$\times$ better energy efficiency and 0.93$\times$ throughput.

\end{itemize}


\section{Background}\label{sec:background}
\subsection{Number Formats}
Number formats play a crucial role in determining the accuracy and performance of a computation. FP and INT are the most common number formats used for ML workloads. We assume that the reader is familiar with these formats.  We present a brief summary of the recently proposed Posit number format that was introduced in 2017~\cite{2017_Gustafson_SFI} to address some of the disadvantages of FP. 

\begin{figure}[h]
\centering{\includegraphics[width=0.8\columnwidth]{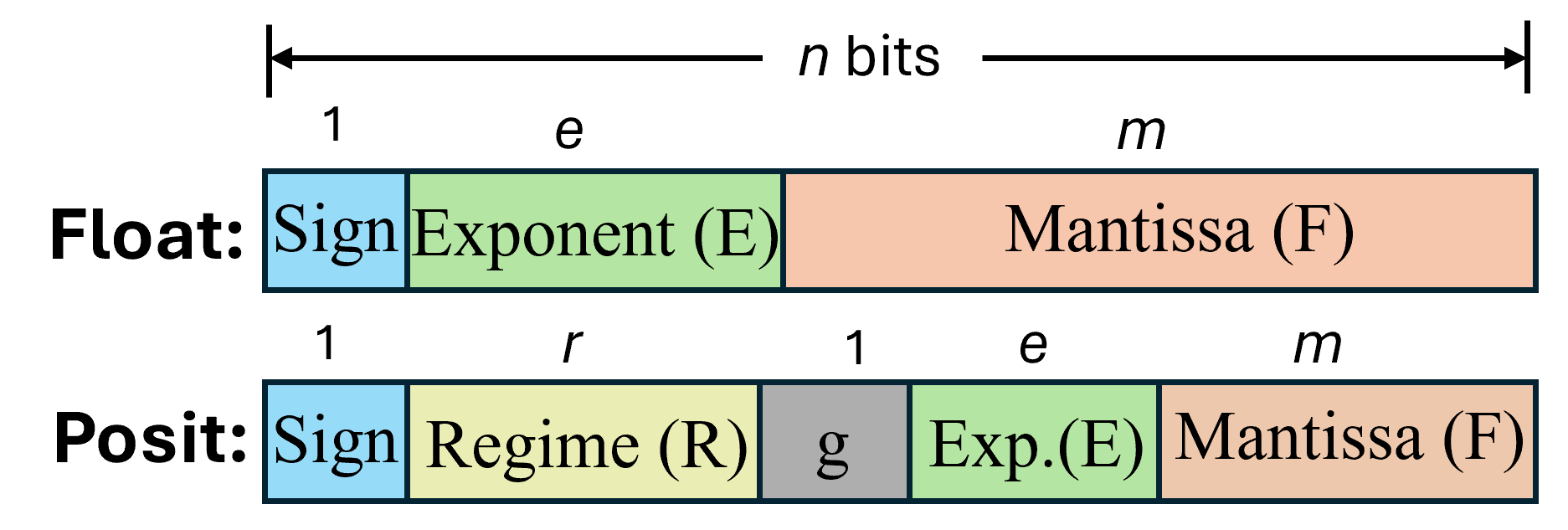}}
\caption{Floating Point and Posit number formats}
\label{Fig:numberformat}
\end{figure}

Fig.~\ref{Fig:numberformat} shows the fields in an FP and a Posit number. In Posit, the lengths of the fields, the \textit{regime}~R (length $r$), the \textit{exponent}~E (length $e$) and the \textit{mantissa}~F (length $m$) are not fixed but vary depending on the number represented. They are only constrained by $r + e + m = n-2$. 

The most important and distinguishing feature of Posit is the regime field $R = (R_{r-1} \cdots, R_0$) of $r$ bits.  The regime field, which is terminated by the stop bit $g = \bar{R_0}$, is either all \underline{1s or all 0s}, and the number of 1s or 0s is used to determine its value denoted by $K$ where $K \in \mathbf{Z}$. If all $r$ bits are 1, then $K = r-1$, where $K \geq 0$, otherwise $K = -r$ where $K < 0$. $K$ is used as an exponent of an exponent. This allows Posit to represent very large and very small numbers more efficiently and with a higher dynamic range, while also minimizing rounding errors. 

Let $P (s, R, g, E, F)$ denote a number in Posit format. Equation~\ref{eq:Positdef} shows the representation of $P$ in decimal $X$. The factor $2^{2^{e}}$ is called \textit{useed (U)}.
\begin{equation}
 \label{eq:Positdef}
    X = 
    \begin{cases}
      0, & \text{if P=}\ 00...0\\
      \infty, & \text{if P=}\ 10...0 \\
      (-1)^{S} \cdot (2^{2^{e} \cdot K}) \cdot 2^{E} \cdot (1+\frac{F}{2^{m}}), & 
     \text{otherwise.}
    \end{cases}
  \end{equation}  

\noindent \textbf{Example of Posit Encoding:} Let $X = 0.00024$, in decimal.  The representation of $X$ in $P(8,2)$ is as follows.  For  $(n = 8, e = 2), U=16$. The regime $R$ is determined by raising $U$ to some integer $K \ni 0 <  \text{U}^K \leq X$. In this case, $K = -3$ $(16^{-3} = 0.00024..., X = 0.00024)$. Then $r = -K = 3$. Therefore, $r = 3$ bits are used to determine $R$ which in this case is $000$ followed by a stop bit $g = 1$. The remaining bits $(n-r-1-e = 3)$ consists of two exponent bits ($e=2$) and the rest ($3-2 =1$) are left for mantissa. Exponent and mantissa are 0 in this case. Therefore, the Posit representation of $0.00024$ is \textcolor{Cyan}{0} \textcolor{Emerald}{0001} \textcolor{Gray}{00} \textcolor{Orange}{0}. The error incurred by Posit encoding of $0.00024$ is 1.6\%. Alternatively, to represent $0.00024$ as an 8-bit floating point number ($(e=3, m=4), (e=4, m=3)$) results in an underflow, i.e. $0.00024$ is rounded to zero, which is represented as \textcolor{Cyan}{0} \textcolor{Emerald}{000} \textcolor{Orange}{0000}. This substantial error can be amplified over multiple operations. For instance, during backpropagation when training a neural network, gradients tend to have very small values which decrease with each iteration. Preserving such values is necessary to achieve high accuracy during training. 

\noindent \textbf{Posit vs. FP:} The Posit number format offers \textit{tapered accuracy}, meaning its precision varies across the representable range, favoring higher accuracy for values near zero~\cite{2017_Gustafson_SFI}. In contrast, FP representation wastes many bit patterns, limiting efficiency. This makes Posit more suitable for computations involving naturally distributed numbers~\cite{2017_Gustafson_SFI}. 


Prior work [17], [18] shows that 16-bit Posit outperforms FP16 in CNN inference accuracy, with 96.54\% vs. 90.65\% on MNIST, and 87.40\% vs. 81.73\% on FashionMNIST. On CIFAR-100, 16-bit Posit matches FP32 accuracy (82.2\%). Additionally, [19] reports that 32-bit Posit achieves mean square error two orders lower than FP32 for 32×32 matrix multiplication over values in [-1, 1], with no performance loss. These results highlight that $n$-bit Posit offers higher precision near zero and a wider dynamic range than $n$-bit FP, benefiting neural network weights and activations which tend to cluster near zero [20], [21]. For more on Posit, see [16], [22].


Despite the flaws, FP is irreplaceable due to Posit compatibility issues with existing systems. In summary, there is no single format that is suitable for all applications. Hence, TC is a viable solution for supporting multiple applications in a system with distinct performance, energy, and accuracy requirements.
\vspace*{.5em}

\noindent Posit hardware implementation challenges:

\begin{itemize}
\item The hardware resources needed to decode a Posit are more than the resources to decode a FP \cite{2024_Jonnalagadda_ESL
}. This is due to the variable length of the fields $(R, E, F)$ in a $n$-bit Posit configuration $P(n,e)$. Whereas, in FP, the fields are fixed in length for a value of $n$, as shown in Fig.~\ref{Fig:numberformat}. Supporting multiple Posit configurations in a system further exacerbates the issue.
 
\item The integration of a Posit compute unit with existing (FP) systems is a challenge due to the overhead cost of decoding hardware, memory management, and controller design required to store both types of data.
\end{itemize}

The proposed work attempts to alleviate some of the above issues as follows:

\begin{itemize}
\item TALU supports different arithmetic and logical operations without using any dedicated hardware units. This increases the flexibility of TALU with a near-zero overhead cost.
 
\item A novel Posit decode algorithm is proposed that uses the same hardware units in TALU as the arithmetic operations. This further reduces the cost of Posit decode as opposed to existing Posit compute units with dedicated decoders. 
\end{itemize}

\section{Transprecision ALU (TALU) Design}\label{sec:PPE}

\subsection{Proposed Posit Computation}
\label{sec:Posit_arithmetic}
To perform operations on Posit numbers, the hardware must extract the fields.  This is the decode operation that is described below. The decode step incurs significant hardware cost as explained 
in~\cite{2024_Jonnalagadda_ESL,2019_Jaiswal_IEEEAccess,2018_Chaurasiya_ICCD,2024_Sun_TCAD}. This is due to the variable-length fields ($R, E, F$) of the Posit number representation. Therefore, dedicated decoder/encoder units are used in the above designs. The increasing value of bitwidth ($n$) further increases the cost of decoder and computation logic. 

In this section a novel Posit decode algorithm is presented to address the overhead cost described above. The proposed algorithm decomposes the decode into a set of operations, listed in Table \ref{tab:Qfn}, that are directly supported by TALU. The decomposition makes the decode algorithm scale with the increasing value of the bitwidth $n$. The hardware overheads are eliminated because, unlike the methods described in existing literature, no dedicated hardware units are allocated. 

The inputs of the Posit decode algorithm are $P$, $n$ and $e$. The algorithm outputs the fields of the representation $P$: $sign (S)$, $regime (R)$, $exponent (E)$, and $mantissa (F)$. Algorithm~\ref{algo:decode} below uses two main functions: $Find\_R$ and $Find\_E\_and\_F$. 

\begin{enumerate}
    \item \textbf{Find\_S}: This function returns the most significant bit (MSB) of the Posit number, i.e., ($P[n-1]$).
    
    \item \textbf{Find\_R}: This function returns the regime value of the Posit $P(n,e)$.  The main idea is to look for the guard bit $g = \bar{R_0}$, shown in Fig.~\ref{Fig:numberformat}. The value of $R$ is determined by Position of $g$. Algorithm Posit\_Decode  performs a set of comparison operations between the binary representation ($P[n-1:0]$) and a set of fixed bit patterns $C$ ($(1,1 \cdots 1), (1,1 \cdots 1,0), \cdots, (1,0, \cdots, 0)$).  This comparison is performed by the \textit{threshold logic} in the compute clusters.  This is a key advantage of the TALU design that is unique. The result of each comparison (1 or 0) is packed into a vector $V_i$ that serves as an address to lookup the regime value.

    \item \textbf{Find\_E\_and\_F}: After we get $R$, $P$ is left-shifted by $R+1$ bits, where $+1$ is for $g$. Thereafter, the E and F bits move towards the most significant bits (MSB). Therefore, extraction of first $e$ bits of $P$ gives the exponent value. The remaining bits are the mantissa bits.

\end{enumerate}

\begin{algorithm}
\small
\caption{Posit\_Decode($P, n, e$)}\label{algo:decode}
\begin{algorithmic}[1]
  \REQUIRE $P$, $n$, $e$
  \ENSURE $S$ (sign), $R$ (regime), $K$ (regime value), $E$ (exponent), $F$ (mantissa)

  \STATE $S \gets P[n-1]$

  \STATE \COMMENT{--- Function Find\_R ---}
  \STATE \textbf{function} Find\_R()
  \STATE \quad $T \gets P$ \textbf{if} $P[n-2]=1$ \textbf{else} $\sim P$
  \FOR{$i = 0$ \TO $n-2$}
    \STATE $V_i \gets 1$ \textbf{if} $T[n-2:0] \ge 2^{\,n-1}-1-(2^i-1)$ \textbf{else} $0$
  \ENDFOR
  \STATE $K \gets LUT[V_i]$
  \STATE $R \gets P[n-2 : n-2-(K+1)]$
  \IF{$P[n-2]=1$}
    \RETURN $(K,R)$
  \ELSE
    \RETURN ($-(K+1),R$)
  \ENDIF
  \STATE \textbf{end function}

  \STATE \COMMENT{--- Function Find\_E\_and\_F ---}
  \STATE \textbf{function} Find\_E\_and\_F($K$)
  \STATE \quad $Q \gets P \ll (K+2)$  \COMMENT{logical left shift}
  \STATE \quad $E \gets Q[n-2 : n-1-e]$
  \STATE \quad $F \gets Q[n-2-(K+2)-e : 0]$
  \RETURN $(E,F)$
  \STATE \textbf{end function}

  \STATE $(K,R) \gets$ Find\_R()
  \STATE $(E,F) \gets$ Find\_E\_and\_F($K$)
  \RETURN $(S,R,K,E,F)$
\end{algorithmic}
\end{algorithm}

\noindent \textbf{Posit Arithmetic}: Posit addition and multiplication algorithms are adopted from~\cite{2018_Chaurasiya_ICCD}. These algorithms are decomposed into a sequence of TALU operations listed in Table~\ref{tab:Qfn} and Table~\ref{tab:Qfn2}. The number of cycles for Posit computations are shown in Table~\ref{tab:MAC_cycle}. Cycles for FP and INT computations are also listed to compare of latency for different data formats executed on TALU.

\subsection{Design of TALU}
\textbf{Threshold Logic and Q-function}: The microarchitecture design of TALU is based on threshold functions. A threshold function $f(x_1, \cdots, x_n)$ is a unate Boolean function whose on-set and off-set are \textit{linearly separable}, i.e. there exists a vector of integer weights $W = (w_1, w_2, \cdots, w_n)$ and a threshold $T$ such that
\vspace{-10 pt}
	\begin{equation}
		\label{eq:ThresholdDefinition}
		f(x_1, x_2, \cdots, x_n) = 1 \Leftrightarrow \sum_{i=1}^{n} w_i x_i \geq T.
	\end{equation}
    
This work implements a more general version of the above threshold function. It is referred to here as a Q-function and is expressed as follows: 
\vspace{-10 pt}
\begin{equation}
\label{eq:Q}
    Q(p,Z_0,X,Z_1,Y) = Z_0 + \sum_{j=0}^{p-1} 2^{j} X_j \geq Z_1 +  \sum_{j=0}^{p-1} 2^{j} Y_j
\end{equation} 

The Q function is implemented for $p=8$, and eight such physical implementations (shown as $Q_0$ to $Q_7$) are used to design the primary and secondary cluster of the TALU, as shown in Fig~\ref{Fig:PE}. The physical implementation of the $Q$-function is a compact sequential circuit block that can perform various arithmetic, logic, and comparison operations as listed in Table~\ref{tab:Qfn} and Table~\ref{tab:Qfn2}, on each clock cycle.

Table~\ref{tab:Qfn} and Table~\ref{tab:Qfn2} show the mapping of the arguments of the Q-function ($Z0, Z1, X$, $Y$) to constant bits and the bits of $p$-bit primary operands A and B, to realize operations listed in these tables. In this paper operand bit-width $p\le8$ is used.

\begin{table*}[ht]
\centering
\caption{\small Arguments to a $Q_i$ function for $p$-bit operations in the Primary Cluster (PC) where $p=8$ and $0 \le i \le p-1$. $\{(p-1)'b0,A_i\}$ is a concatenated bit-string with $p-1$ 0's appended to $A_i$.}
\label{tab:Qfn}
\resizebox{1.5\columnwidth}{!}
{\begin{tabular}{|c|c|c|c|c|c|}

\hline
\textbf{Operation}     & \textbf{$Z_0$}     & \textbf{$X$}                &\textbf{$Z_1$}   &\textbf{$Y$}     &\textbf{Output}                                          \\ \hline
AND  & 0 & $\{(p-1)'b0,A_i\}$ & 1 & $\{(p-1)'b0,\sim B_i\}$ & $A_i \wedge B_i$                                      \\ \hline
OR             & 0      & $\{(p-1)'b0,A_i\}$               & 0    & $\{(p-1)'b0,\sim B_i\}$   & $A_i \vee B_i$                                    \\ \hline
NOT            & 0      & $\{(p-1)'b0,\sim B_i\}$               & 1    & 0  & $\sim B_i$                                           \\ \hline
COMP          & 0     & $\{(p-i-1)'b0,A{[}i:0{]}\}$     & 0    & $\{(p-i-1)'b0,B{[}i:0{]}\}$   & $A[i:0] \geq B[i:0]$                                \\ \hline
ADD (Step 1: Carry)  & $C_0$ & $\{(p-i-1)'b0,A{[}i:0{]}\}$ & 1 & $\{(p-i-1)'b0,\sim B{[}i:0{]}\}$ & $Carry_{i+1}$ \\ \hline

XOR (Step 1)          & 0     & $\{(p-1)'b0,A_i\}$      & 1    & $\{(p-1)'b0,\sim B_i\}$        & $A_i \wedge B_i$  ($AND_i$)                     \\ \hline

Posit Decode   & 0     & $P(p,e)$      & 0    &  $2^{p-1}-1-(2^i-1)$        & $V_i$                       \\ \hline
\end{tabular}
}
\end{table*}

\begin{table*}[ht]
\centering
\caption{\small Arguments to a $Q_i$ function for $p$-bit operations in the Secondary Cluster (SC)}
\label{tab:Qfn2}
\resizebox{1.5\columnwidth}{!}
{\begin{tabular}{|c|c|c|c|c|c|}

\hline
\textbf{Operation}     & \textbf{$Z_0$}     & \textbf{$X$}                &\textbf{$Z_1$}   &\textbf{$Y$}  &\textbf{Output}                                            \\ \hline

ADD (Step 2: Sum)  & $A_i$ & $\{(p-1)'b0,B_i\}$ & 0 & $ \{(p-2)'b0, Carry_{i+1}, \sim Carry_{i}\}$ & $A_i+B_i$\\ \hline
XOR (Step 2)          & $A_i$     & $\{(p-1)'b0,B_i\}$  & 1    & $\{(p-2)'b0, AND_i, 0\}$     & $A_i \oplus B_i$                               \\ \hline
\end{tabular}
}
\end{table*}

The TALU design is shown in Fig.~\ref{Fig:PE}, and a detailed description is given in Section~\ref{subsec:TALU_Microarchitecture} below. An array of $p$ independent sequential blocks can produce a $p$-bit output in a clock cycle. A notable difference between Table~\ref{tab:Qfn} and Table~\ref{tab:Qfn2}, is that ADD, XOR/XNOR require two cycles (XOR/XNOR and ADD require two threshold functions) and other functions require one cycle. Therefore, Table~\ref{tab:Qfn2} only has XOR and ADD functions which shows how the second cluster is used to complete the addition and exclusive-or operations- Step 2, after the first cluster computes Step 1, shown in Table~\ref{tab:Qfn}. Addition is performed using carry-lookahead. While the carry-out is a threshold function of its inputs, the sum is not.  However the sum can also be expressed as a threshold function of its inputs \underline{and its carry-out}. This is the key merit of the using threshold functions and the Q-function as a template. More on Q-function can be referenced from the work \cite{2022_QNN_wagle}.

\subsection{TALU Microarchitecture}
\label{subsec:TALU_Microarchitecture}

\begin{figure}[]
\centering{\includegraphics[width=1\columnwidth]{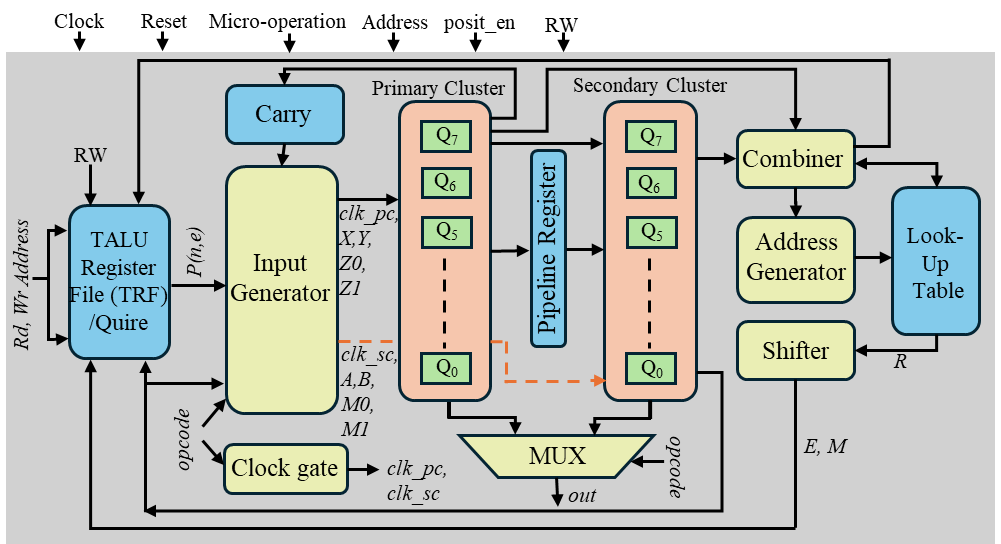}}
\caption{Proposed Transprecision Arithmetic and Logic Unit (TALU) design. For physical implementation of Q-function, $p=8$.}
\label{Fig:PE}
\end{figure}

The main modules in TALU are the compute clusters: Primary Cluster (PC) and Secondary cluster (SC). These identical clusters serve as the core of TALU and can be operated in sequence or in parallel. \textit{The unique cluster design enables them to perform all the operations listed in Table~\ref{tab:Qfn}}. The clusters consist of sequential blocks that implement independent Q functions (marked as $Q_i$ in Fig.~\ref{Fig:PE}) where $0\le i \le 7$. For an $n=8$ bit Posit decode, only one cluster (PC) is used and $p=8$, whereas for $n=16$ bit Posit decode, both the PC and SC operate in comcurrently each with $p=8$. For Posit addition/subtraction operations, the clusters operate in a pipelined manner: the PC generates the carry bits, which are then transferred to the SC via a pipeline register to compute the final result in the next cycle.  

The format (Posit or FP/INT) is determined by the primary control signal posit\_en shown in Fig.~\ref{Fig:PE}. When enabled, TALU is configured to perform Posit operations and FP/INT operations otherwise. TALU is configured to perform the operations in Table~\ref{tab:Qfn} through the issuance of micro-operations. The input signal RW is 1 or 0 for a Read or Write operation, respectively. Operands are accessed from the TALU Register File (TRF). The primary operands are fed into the Input Generator, producing the arguments $Z_0, X, Z_1, Y$ for the clusters depending upon the operation.

\noindent \textbf{Mapping of the Algorithm~\ref{algo:decode} on TALU}: The main function of the Algorithm~1 is Find\_R. It consists of multiple comparison operations between the Posit ($T[n-2:0]$) and the precomputed constants ($2^{p-1}-1-(2^{i}-1)$) where $p=n, 0<i\le n-2$. These are mapped to the Q-function implemented by the sequential blocks, each producing one bit. To decode an 8-bit $P(n,e)$, a total of seven Q-functions are needed and can be computed concurrently by the PC. The boxed operation in Algorithm \ref{algo:decode} is the step that is mapped on the PC's Q-functions. Thereafter, the output bits $V_i$ are directed to the Address Generator, via the Combiner, to produce the address to the \text{Look-Up Table (LUT)} containing precomputed regime values ($K$), indicative of $R$. By transforming $P(n,e)$ into $T$ (as described in Algorithm~1), the number of possible $R$ values ranges from $[0,6]$. Therefore, the size of LUT is only a few bits, and the \underline{LUT size is the same} for decoding 8 and 16 bit Posit numbers. Combiner is used to find regime by logically combining the outputs of both the clusters receiving the MSB and LSB bits of 16-bit Posit. The regime value is passed to the Shifter to produce the exponent and mantissa bits ($E, F$), as described in \ref{algo:decode}. The $K, E, F$ values are subsequently stored back into the TRF. 

For example, let $P(8,2) = 01110100$ be a Posit representation of a number. To decode $P$, $P[n-2:0]$ are mapped to input $X$, 0 to $Z_0$ and $Z_1$ and precomputed constants ($2^{p-1}-1-(2^{i}-1)$) are mapped to $Y$  of each of the eight $Q_i$ units in the PC (where $p=n=8$). The output bits $V_i$ from each Q-function are evaluated as $\{V_6, V_5, \cdots, V_0\}= \{0,0,0,0,0,1,1,1\}$ which is used as an index to the LUT to get $K=2$. The useed $U = 2^{2^e}$ is raised to $K$ in the Posit definition (see Equation~\ref{eq:Positdef}). $K$ is used to extract the regime field of $P$ where $R = P[p-2:(p-2)-(K+1)]= 1110$. To determiend $E$ and $M$, $P[p-2:0]$ is left shifted by $K + 2$ resulting in  $1000000$ as the shifted value of $P$.  The first $e = 2$ bits $10$ are $E$. The bits of $M$ are the following $(p-2-e)-(K+2):0$ bits of P i.e., $P[6-4-2:0]=0$. Thus $P$ represents the decimal number $16^2 * 2=512$ as per the Posit definition in equation \ref{eq:Positdef}.

For 16-bit decode, both clusters perform the same set of comparison operations concurrently, and their respective outputs are looked up sequentially, producing two regime values, which are then logically combined using the Combiner module. Therefore, each compute cluster consists of seven Q-functions to minimize the delay of the decode algorithm by maximizing the number of parallel operations. Decoding two 8-bit Posits or a 16-bit Posit requires two clock cycles in TALU, regardless of the $e$ value. Once the decoded Posit fields $S, R, E, F$ for the operands are computed, they are stored in TRF. They are retrieved from the TRF in the subsequent cycle to perform the intended operation. Note that TALU is a two-stage pipelined design that can perform ALU operations as opposed to a specialized Multiply-Accumulate Unit (MAC).

\subsection{Transprecision Vector Processor Unit (TALU-V)}\label{TALU-Vrisc}

The most common ML compute kernel is matrix multiplication (MATMUL)~\cite{2022_Chander_VLSID}. The kernels can be decomposed into a sequence of SIMD vector operations. In order to operate on two $N$-sized vector operands, $N$ TALUs are used in a transprecision vector processor unit (TALU-V). TALU-V is interfaced with the $32 \times 32$ register file of a low-power RISC-V processor called RISCY~\cite{2017_Schiavone_PATMOS}. Consequently, the RISCY instruction set can be extended for custom vector instructions that run on the integrated TALU-V rather than the native RISCY scalar ALU. 

\begin{figure}[htb]
\centering{\includegraphics[width=0.8\columnwidth]{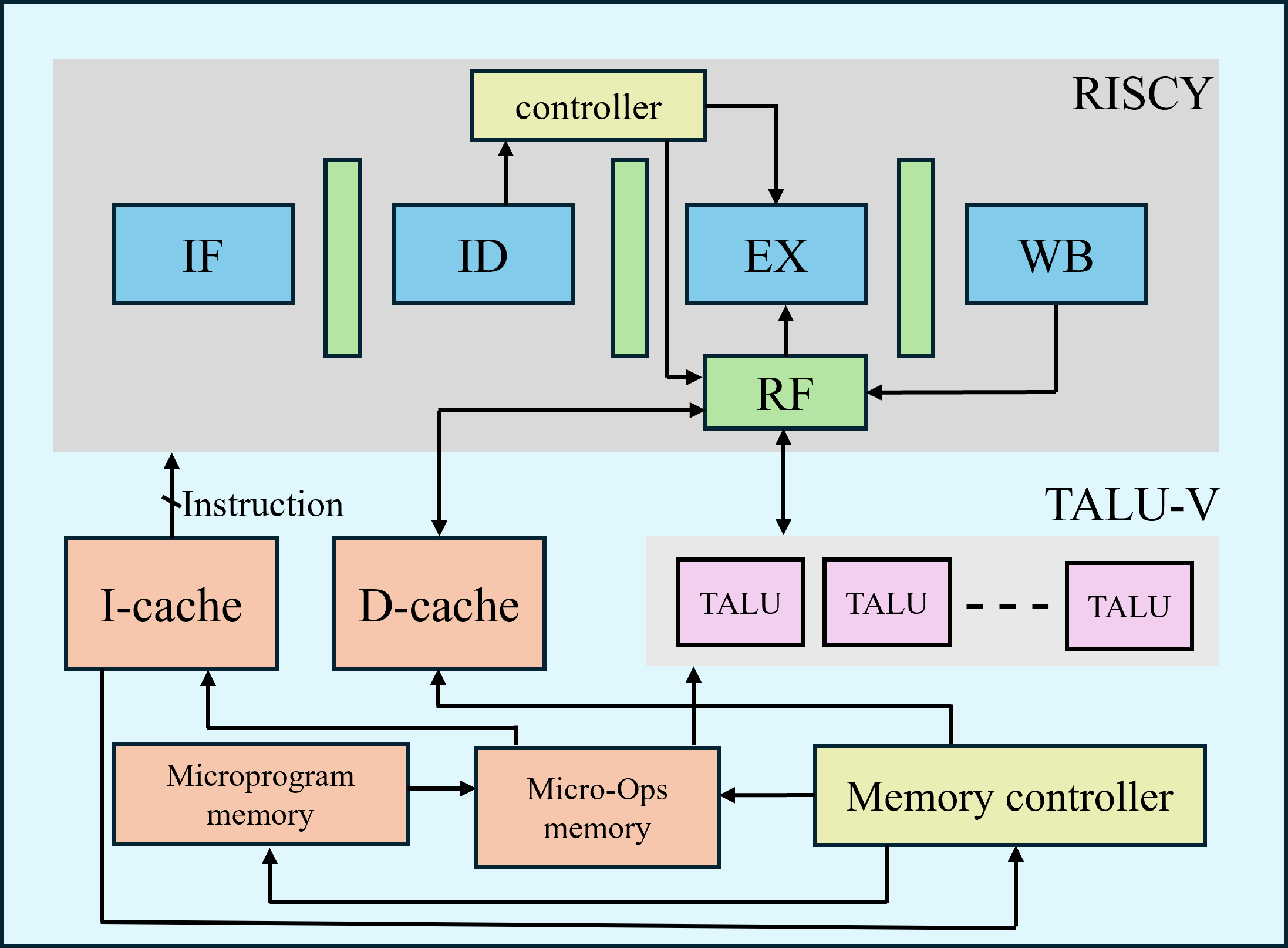}}
\caption{Transprecision Vector Processor Unit integrated to RISCY\cite{2017_Gautschi_TVLSI} (RISCY+TALU-V) }
\label{Fig:RISC-V}
\end{figure}

An overview of the whole architecture (RISCY+TALU-V) is shown in Fig. \ref{Fig:RISC-V}. The details of the architecture and RISCY ISA extension for vector operations is a part of future work. The aim of this work is to design a compute element for applications on a smart edge processor with ultra-low power requirements.

\vspace{-10 pt}
\section{Experiments and Results}\label{Exp}

\subsection{Methodology}\label{subsec:setup}

A Verilog description of TALU, using Equation~\ref{eq:Q} as the functional description of the Q-function sequential block was synthesized at 2 GHz using Cadence Genus with a STM 28nm library and placed and routed using Cadence Innovus. A Python-based cycle-level simulator was used for estimating the number of cycles for Posit computations, shown in Table~\ref{tab:MAC_cycle}. Posit computation is verified by running the same computations on an open-source library~\textit{softposit}~\cite{2020_Klower_SoftPosit}. We first note that TALU is a bit-sliced design and hence the number of cycles it takes to perform an INT, FP or Posit operation increased with the bitwidth ($n$). 

\vspace{-10 pt}
\begin{table}[h]
\centering
\caption{Number of clock cycles for Multiplication and Addition for TALU for different Posit configurations $(n, e)$ and standard Floats}
\label{tab:MAC_cycle}
\begin{tabular}{|c|c|c|c|}
\hline
Configuration & \begin{tabular}[c]{@{}c@{}}Decode \\ cycles\end{tabular} & \begin{tabular}[c]{@{}c@{}}Multiplication\\ cycles\end{tabular} & \begin{tabular}[c]{@{}c@{}}Addition \\ cycles\end{tabular} \\ \hline
P(8,0)        & 2                                                        & 17                                                              & 21                                                         \\ \hline
P(8,2)        & 2                                                        & 19                                                              & 23                                                         \\ \hline
P(16,0)       & 6                                                        & 25                                                              & 23                                                         \\ \hline
P(16,2)       & 6                                                        & 29                                                              & 25                                                         \\ \hline

FP8           & 0                                                        & 18                                                              & 8                                                          \\ \hline
FP16          & 0                                                        & 87                                                              & 10                                                          \\ \hline
INT4          & 0                                                        & 13                                                               & 2                                                          \\ \hline
INT8          & 0                                                        &  28                                                             &  2                                                         \\ \hline
INT16         & 0                                                        & 105                                                             & 4                                                          \\ \hline

\end{tabular}
\end{table}

\subsection{Comparison against State-of-The-Art Posit only Designs}

TALU is compared with three state-of-the-art designs: (1)~DFMA~\cite{2020_Neves_SiPS}, (2)~VMULT~\cite{2021_Zhang_ISCAS}, and (3)~Fused MAC with Kogge-Stone Adder~\cite{2021_Murillo_ICCD} that compute MAC operations in Posit format only. 
These MAC units operate on 8, 16, and 32-bit Posit with $e=2$\footnote{$e=2$ is a standard value used in hardware designs~\cite{2020_Guntoro_DATE}}. The results are shown in Table~\ref{tab:Posit_only}. The DFMA and Fused MAC designs have \underline{separate} implementations for different bitwidths (8, 16, 32 bits). Therefore, unlike TALU and VMULT, DFMA and Fused MAC cannot be reconfigured to change the bitwidth of the computation at runtime. Hence, DFMA and Fused MAC cannot perform TC.

\def\twoStack#1#2{\begin{tabular}[c]{@{}c@{}}#1\\ #2\end{tabular}}
\def\threeStack#1#2#3{\begin{tabular}[c]{@{}c@{}@{}} #1\\#2\\#3\end{tabular}}
\def\fourStack#1#2#3#4{\begin{tabular}[c]{@{}c@{}@{}@{}} #1\\#2\\#3\\#4\end{tabular}}

\begin{table}[h]
\caption{Comparison of the proposed TALU (placed and routed) with {\color{red} Posit-only compute elements scaled to 28 nm technology}}
\label{tab:Posit_only}
\resizebox{\columnwidth}{!}
{\begin{tabular}{|c|c|c|c|c|c|c|c|}
\hline
\twoStack{Compute}{Element} &
\#bits &
\twoStack{Delay}{(ns)} &
\twoStack{Area}{($mm^2$)} &
\twoStack{Power}{(mW)} & 
\twoStack{PDP}{(pJ)} &
\twoStack{Pow. Den.}{(mW/$mm^2)$} \\ \hline
%
%
\fourStack{TALU}{$\approx$ 2GHz}{28nm, P\&R}{\color{red} Posit, FP and INT} &
\threeStack{8}{16}{32} &
\threeStack{21.5}{24}{25.5} &
0.0026 &
1.81 &
\threeStack{38.9}{43.44}{46.15} &
696.15 \\ \hline
%
%
\fourStack{VMULT~\cite{2021_Zhang_ISCAS}}{$\approx$ 400 MHz}{90nm, Synth.} {\color{red} Posit Only}&
\threeStack{8}{16}{32} &
0.71   & 
0.014 &
42.94 &
30.7 &
2878.62  
 \\ \hline
%
%
\fourStack{DFMA~\cite{2020_Neves_SiPS}}{$\approx$ 800MHz}{45nm, Synth.} {\color{red} Posit Only} &
\threeStack{8}{16}{32} &
\threeStack {0.75} {0.93} {1.12} & 

\threeStack{0.0044} {0.0145} {0.0435} & 

\threeStack{13.77} {32.4} {76.95} & 

\threeStack{10.28} {30.24} {86.18} & 

\threeStack{3155} {2227.5} {1767.1}  \\ \hline
%
%
\fourStack{Fused MAC~\cite{2021_Murillo_ICCD}}{Kogge-Stone Adder}{$\approx$ 1 GHz, 45nm Synth.} {\color{red} Posit Only} &
\threeStack{8}{16}{32} & 

\threeStack{0.50} {0.47} {0.63} &

\threeStack{0.0023} {0.006} {0.015} &

\threeStack{3.92} {9.5} {27.44} & 

\threeStack{1.97} {4.55} {17.41} &

\threeStack{1724.97} { 1609.28} {1829.52} 
\\ \hline
\end{tabular}}
\end{table}

TALU is designed in 28nm, whereas DFMA and Fused MAC were designed in 45nm, and VMULT was designed in the 90nm technology node. For the purpose of comparison, values of DFMA, VMULT, and Fused MAC are scaled to 28nm~\cite{2017_Stillmaker_scaling}. Note that, unlike TALU, the numbers associated with the other designs are based on synthesis results, which are less accurate.

\def\twoStack#1#2{\begin{tabular}[c]{@{}c@{}}#1\\ #2\end{tabular}}
\def\threeStack#1#2#3{\begin{tabular}[c]{@{}c@{}@{}} #1\\#2\\#3\end{tabular}}
\def\fourStack#1#2#3#4{\begin{tabular}[c]{@{}c@{}@{}@{}} #1\\#2\\#3\\#4\end{tabular}}

\noindent \textbf{Key takeaways from Table~\ref{tab:Posit_only}}: (\textbf{1})~TALU, being a bit-sliced design, has a substantially higher delay $\approx 22X~to~50X$ compared to the other single-cycle designs. 
(\textbf{2})~The trade-off is a substantial reduction in area of $\approx 2X~to~17X$ and power of $\approx 2X~to~43X$. (\textbf{3})~The power density (watts/area) of TALU is $\approx 2X~to~5X$ lower than the other designs. \textbf{(4)} Energy (PDP) for TALU is $\approx 1.5X~to~20X$ worse than the other designs. This is because TALU supports Posit, FP and INT computations, whereas the other designs only support Posit computation.

The reduction in power, area, and power density are the most important metrics for the intended  application domain such as wearables, handheld medical devices and many other battery-powered, small form-factor mobile systems. Table~\ref{tab:Posit_only} shows that TALU outperforms the other designs by substantial factors. 

\subsection{Comparison of TALU with UMAC~\cite{2022_Crespo_IEEETCS_UM}}

For a fairer comparison, we compare TALU with the Unified MAC unit (UMAC) presented in~\cite{2022_Crespo_IEEETCS_UM}, which supports 8, 16, and 32 bit Posit and FP computations in a single unit. Similar to TALU, the UMAC can also be configured at runtime to change the bitwidth and number format. However, TALU supports Posit, FP and INT, whereas UMAC supports Posit and FP only. Nonetheless, UMAC comes closest to TALU in functionality among other compute elements. UMAC is a six-stage pipelined computing element, \textit{synthesized} using 28-nm UMC technology at 667 MHz, with the output of the size 32$\times$1$/$16$\times$2$/$8$\times$4 produced per cycle. 

\vspace{-10 pt}
\begin{table}[h]
\caption{Comparison of the proposed TALU (Posit, FP and INT) with Unified (Posit and FP) MAC Unit at 28 nm technology}

\resizebox{\columnwidth}{!}
{\begin{tabular}{|c|c|c|c|c|c|c|c|}
\hline
\twoStack{Compute}{Element} &
\#bits &
\twoStack{Delay}{(ns)} &
\twoStack{Area}{($mm^2$)} &
\twoStack{Power}{(mW)} & 
\twoStack{PDP}{(pJ)} &
\twoStack{Pow. Den.}{(mW/$mm^2)$} \\ \hline
%
%
\threeStack{TALU}{$\approx$ 2GHz}{28nm, P\&R} &
\threeStack{8}{16}{32} &
\threeStack{21.5}{24}{25.5} &
0.0026 &
1.81 &
\threeStack{38.9}{43.44}{46.15} &
696.15 \\ \hline
%
\threeStack{UMAC}{$\approx$ 667 MHz}{28nm, Synth.} &
\threeStack{8}{16}{32} &
1.5 &
\twoStack {0.0515}{ (\textbf{\color{teal} 19.8$\times$})} &
\twoStack {99} { (\textbf{\color{teal} 54.6$\times$})}  &
\twoStack{148.50}{ (\textbf{\color{teal} 3.47$\times$})}  &
\twoStack{1941.17}{ (\textbf{\color{teal} 2.76$\times$})}  \\ \hline
\end{tabular}}
\label{tab:UMcomp}
\end{table}

Table~\ref{tab:UMcomp} shows the comparison of TALU and UMAC processing elements. TALU is \textbf{20$\times$} smaller and consumes \textbf{54.6$\times$} lower power than UMAC. The larger area of the UMAC is due to the significant area of the decode/encode modules for Posit/FP and a large booth multiplier. In contrast, the proposed TALU has no dedicated encoders, decoders or multiplier units. Although both the power and area of TALU are individually smaller, the \textbf{$ \approx 3\times$} lower power density of TALU compared to the UMAC implies that the power reduced much more than area. The energy consumption of a TALU for a MAC operation is \textbf{$ \approx 3.5\times$} lower than energy consumption of a UMAC for the same operation. This shows that TALU is best suited for battery-powered, small form-factor mobile devices.

\subsection{Executing ML compute kernels on the RISCY+TALU-V}
\label{subsec:RISC-V}

The RISCY+TALU-V architecture, described in Section \ref{TALU-Vrisc}, is evaluated for energy efficiency and throughput while running $3 \times 3$ matrix  multiplications (MATMUL), as they are the dominant ML compute kernels for DNNs. A MATMUL operation can be scheduled on the TALU-V vector processor as vector operations. In this work, Posit $P(8,2)$ is exclusively used for vector operations, as this configuration is most used for DNNs deployed on edge devices~\cite{2018_Carmichael_ArXiv}. However, if an application requires FP/INT vector computation, then RISCY+TALU-V design can be switched to perform FP/INT computations without any performance overhead.

Several RISC-V-based vector processors capable of tensor computations have been proposed in the literature~\cite{2019_Mach_VLSI-SOC,2024_Bertaccini_IEEE}. However, these processors only support FP formats such as FP8, FP16, FP32, Bfloat, and INT data types, without any support for Posit arithmetic. To enable a direct comparison between the proposed RISCY+TALU-V architecture and another RISCY-based vector processor capable of both Posit and FP arithmetic, we use the vector unit of the Unified MAC (UMAC) processor presented in~\cite{2022_Crespo_IEEETCS_UM}. Similar to TALU-V, the UMAC vector unit (UMAC-V) is integrated with the RISCY CPU to form the RISCY+UMAC-V architecture. 

For a fair comparison between the RISCY+TALU-V and RISCY+UMAC-V architectures, an \textit{equi-area} analysis is performed. The area of TALU-V and UMAC-V vector processors are matched, assuming the same RISCY  is used in both architectures. The register file (RF) can provide 1024 bits in parallel to the vector processors. Each TALU unit is designed to take 8 bits as input. Therefore, 128 TALUs are used to access 1024 bits of the RF in parallel. The UMAC is $ \approx 20\times$ larger than TALU and is designed to take 96 bits as input. Therefore, to match the area of 128 TALUs, 6 UMAC units are used with parallel access of 576 bits of RF. The higher latency of TALU against a UMAC is compensated by having more TALUs operating in parallel in the TALU-V compared to the number of UMAC units in the UMAC-V.

The compact design of TALU is exploited to make it operate at a much higher frequency. TALU in 28nm achieved timing closure during P\&R at 2~GHz. The reported maximum frequency for UMAC was 667~MHz. Table~\ref{tab:architecture_comp} demonstrates the above points, showing that TALU achieved nearly the same throughput (number of $3 \times 3$ matrix multiplications/sec) as UMAC, but was nearly twice as energy efficient. 

\vspace{-10 pt}
\begin{table}[h]
\centering
\caption{Ratio of TALU-V+RISCY versus UMAC-V+RISCY architectures for energy efficiency (\#Kernels/J) and throughput (\#Kernels/s) for 3X3 matrix multiplication kernel}
\begin{tabular}{|c|cc|}
\hline
\multirow{2}{*}{Kernel} & \multicolumn{2}{c|}{Equi-area Comparison}                    \\ \cline{2-3} 
                                 & \multicolumn{1}{c|}{Throughput} & Energy Efficiency \\ \hline
MATMUL                           & \multicolumn{1}{c|}{0.93x}               & 1.98x                      \\ \hline
\end{tabular}
\label{tab:architecture_comp}
\end{table}

\vspace{-10 pt}
\section{Conclusion}
\label{sec:conclusion} 
This work proposes a highly energy-efficient compute element, TALU, that is suitable for edge devices with ultra-low power and area requirements. The uniqueness of TALU is due to use of transprecision computing, custom designed mathematical macros called Q function to realize diverse functionality and performing decode and arithmetic operations as a sequence of such Q functions that eliminate the need for any dedicated decode/encode units. This paper also presented the design of vector unit consisting of multiple TALUs integrated with a low power RISC-V based architecture, to perform vector operations with high energy efficiency.



\bibliographystyle{IEEEtran}
\bibliography{references}

\begin{thebibliography}{10}
\providecommand{\url}[1]{#1}
\csname url@samestyle\endcsname
\providecommand{\newblock}{\relax}
\providecommand{\bibinfo}[2]{#2}
\providecommand{\BIBentrySTDinterwordspacing}{\spaceskip=0pt\relax}
\providecommand{\BIBentryALTinterwordstretchfactor}{4}
\providecommand{\BIBentryALTinterwordspacing}{\spaceskip=\fontdimen2\font plus
\BIBentryALTinterwordstretchfactor\fontdimen3\font minus \fontdimen4\font\relax}
\providecommand{\BIBforeignlanguage}[2]{{%
\expandafter\ifx\csname l@#1\endcsname\relax
\typeout{** WARNING: IEEEtran.bst: No hyphenation pattern has been}%
\typeout{** loaded for the language `#1'. Using the pattern for}%
\typeout{** the default language instead.}%
\else
\language=\csname l@#1\endcsname
\fi
#2}}
\providecommand{\BIBdecl}{\relax}
\BIBdecl

\bibitem{2022_Crespo_IEEETCS_UM}
L.~Crespo, P.~Tomás, N.~Roma, and N.~Neves, ``Unified posit/ieee-754 vector mac unit for transprecision computing,'' \emph{IEEE Transactions on Circuits and Systems II: Express Briefs}, vol.~69, no.~5, pp. 2478--2482, 2022.

\bibitem{2021_Mach_TVLSI}
S.~Mach, F.~Schuiki, F.~Zaruba, and L.~Benini, ``Fpnew: An open-source multiformat floating-point unit architecture for energy-proportional transprecision computing,'' \emph{IEEE Transactions on Very Large Scale Integration (VLSI) Systems}, vol.~29, no.~4, pp. 774--787, 2021.

\bibitem{2018_Malossi_DATE}
A.~C.~I. Malossi, M.~Schaffner, A.~Molnos, L.~Gammaitoni, G.~Tagliavini, A.~Emerson, A.~Tomás, D.~S. Nikolopoulos, E.~Flamand, and N.~Wehn, ``The transprecision computing paradigm: Concept, design, and applications,'' in \emph{2018 Design, Automation \& Test in Europe Conference \& Exhibition (DATE)}, 2018, pp. 1105--1110.

\bibitem{2020_Grosser_ACM}
T.~Grosser, T.~Theodoridis, M.~Falkenstein, A.~Pitchanathan, M.~Kruse, M.~Rigger, Z.~Su, and T.~Hoefler, ``Fast linear programming through transprecision computing on small and sparse data,'' \emph{Proceedings of the ACM on Programming Languages}, vol.~4, pp. 1--28, 11 2020.

\bibitem{2018_Chen_CryptographyBook}
H.~Chen, K.~Laine, R.~Player, and X.~Yuhou, \emph{High-Precision Arithmetic in Homomorphic Encryption}, 03 2018, pp. 116--136.

\bibitem{2023_Leong_Springer}
\BIBentryALTinterwordspacing
S.~H. Leong and J.~L. Gustafson, ``Lossless ffts using posit arithmetic,'' in \emph{Next Generation Arithmetic: 4th International Conference, CoNGA 2023, Singapore, March 1-2, 2023, Proceedings}.\hskip 1em plus 0.5em minus 0.4em\relax Berlin, Heidelberg: Springer-Verlag, 2023, p. 1–18. [Online]. Available: \url{https://doi.org/10.1007/978-3-031-32180-1_1}
\BIBentrySTDinterwordspacing

\bibitem{2021_Romanov_IEEEAccess}
A.~Y. Romanov, A.~L. Stempkovsky, I.~V. Lariushkin, G.~E. Novoselov, R.~A. Solovyev, V.~A. Starykh, I.~I. Romanova, D.~V. Telpukhov, and I.~A. Mkrtchan, ``Analysis of posit and bfloat arithmetic of real numbers for machine learning,'' \emph{IEEE Access}, vol.~9, pp. 82\,318--82\,324, 2021.

\bibitem{2022_Dube_ICCAD}
A.~Dube, A.~Wagle, G.~Singh, and S.~Vrudhula, ``Tunable precision control for approximate image filtering in an in-memory architecture with embedded neurons,'' in \emph{2022 IEEE/ACM International Conference On Computer Aided Design (ICCAD)}, 2022, pp. 1--9.

\bibitem{2023_Baleen_ArXiv}
\BIBentryALTinterwordspacing
M.~van Baalen, A.~Kuzmin, S.~S. Nair, Y.~Ren, E.~Mahurin, C.~Patel, S.~Subramanian, S.~Lee, M.~Nagel, J.~B. Soriaga, and T.~Blankevoort, ``Fp8 versus int8 for efficient deep learning inference,'' \emph{ArXiv}, vol. abs/2303.17951, 2023. [Online]. Available: \url{https://api.semanticscholar.org/CorpusID:257900777}
\BIBentrySTDinterwordspacing

\bibitem{2024_SHo_DeepSeek}
\BIBentryALTinterwordspacing
Z.~Shao, D.~Dai, D.~Guo, B.~L.~B. Liu), Z.~Wang, and H.~Xin, ``Deepseek-v2: A strong, economical, and efficient mixture-of-experts language model,'' \emph{ArXiv}, vol. abs/2405.04434, 2024. [Online]. Available: \url{https://api.semanticscholar.org/CorpusID:269613809}
\BIBentrySTDinterwordspacing

\bibitem{2017_Gautschi_TVLSI}
M.~Gautschi, P.~D. Schiavone, A.~Traber, I.~Loi, A.~Pullini, D.~Rossi, E.~Flamand, F.~K. Gürkaynak, and L.~Benini, ``Near-threshold risc-v core with dsp extensions for scalable iot endpoint devices,'' \emph{IEEE Transactions on Very Large Scale Integration (VLSI) Systems}, vol.~25, no.~10, pp. 2700--2713, 2017.

\bibitem{2019_Zhang_TC}
H.~Zhang, D.~Chen, and S.-B. Ko, ``Efficient multiple-precision floating-point fused multiply-add with mixed-precision support,'' \emph{IEEE Transactions on Computers}, vol.~68, no.~7, pp. 1035--1048, 2019.

\bibitem{2021_Zhang_ISCAS}
H.~Zhang and S.-B. Ko, ``Efficient multiple-precision posit multiplier,'' in \emph{2021 IEEE International Symposium on Circuits and Systems (ISCAS)}, 2021, pp. 1--5.

\bibitem{2020_Neves_SiPS}
N.~Neves, P.~Tomás, and N.~Roma, ``Dynamic fused multiply-accumulate posit unit with variable exponent size for low-precision dsp applications,'' in \emph{2020 IEEE Workshop on Signal Processing Systems (SiPS)}, 2020, pp. 1--6.

\bibitem{2017_Gustafson_SFI}
\BIBentryALTinterwordspacing
Gustafson and Yonemoto, ``Beating floating point at its own game: Posit arithmetic,'' \emph{Supercomput. Front. Innov.: Int. J.}, vol.~4, no.~2, p. 71–86, Jun. 2017. [Online]. Available: \url{https://doi.org/10.14529/jsfi170206}
\BIBentrySTDinterwordspacing

\bibitem{2024_Jonnalagadda_ESL}
A.~A. Jonnalagadda, R.~Thotli, S.~Veeramachaneni, U.~A. Kumar, and S.~E. Ahmed, ``Energy-efficient decoding and encoding hardware for optimized posit arithmetic,'' \emph{IEEE Embedded Systems Letters}, pp. 1--1, 2024.

\bibitem{2019_Jaiswal_IEEEAccess}
M.~K. Jaiswal and H.~K.-H. So, ``Pacogen: A hardware posit arithmetic core generator,'' \emph{IEEE Access}, vol.~7, pp. 74\,586--74\,601, 2019.

\bibitem{2018_Chaurasiya_ICCD}
R.~Chaurasiya, J.~Gustafson, R.~Shrestha, J.~Neudorfer, S.~Nambiar, K.~Niyogi, F.~Merchant, and R.~Leupers, ``Parameterized posit arithmetic hardware generator,'' in \emph{2018 IEEE 36th International Conference on Computer Design (ICCD)}, 2018, pp. 334--341.

\bibitem{2024_Sun_TCAD}
J.~Sun and Y.~Lao, ``Efficient data extraction circuit for posit number system: Ldd-based posit decoder,'' \emph{IEEE Transactions on Computer-Aided Design of Integrated Circuits and Systems}, vol.~43, no.~6, pp. 1919--1923, 2024.

\bibitem{2022_QNN_wagle}
A.~Wagle, G.~Singh, S.~Khatri, and S.~Vrudhula, ``An asic accelerator for qnn with variable precision and tunable energy efficiency,'' \emph{IEEE Transactions on Computer-Aided Design of Integrated Circuits and Systems}, vol.~43, no.~7, pp. 2057--2070, 2024.

\bibitem{2022_Chander_VLSID}
V.~N. Chander and K.~Varghese, ``A soft risc-v vector processor for edge-ai,'' in \emph{2022 35th International Conference on VLSI Design and 2022 21st International Conference on Embedded Systems (VLSID)}, 2022, pp. 263--268.

\bibitem{2017_Schiavone_PATMOS}
P.~Davide~Schiavone, F.~Conti, D.~Rossi, M.~Gautschi, A.~Pullini, E.~Flamand, and L.~Benini, ``Slow and steady wins the race? a comparison of ultra-low-power risc-v cores for internet-of-things applications,'' in \emph{2017 27th International Symposium on Power and Timing Modeling, Optimization and Simulation (PATMOS)}, 2017, pp. 1--8.

\bibitem{2020_Klower_SoftPosit}
M.~Klöwer, P.~Düben, and T.~Palmer, ``Number formats, error mitigation, and scope for 16‐bit arithmetics in weather and climate modeling analyzed with a shallow water model,'' \emph{Journal of Advances in Modeling Earth Systems}, vol.~12, 10 2020.

\bibitem{2021_Murillo_ICCD}
R.~Murillo, D.~Mallasén, A.~A. Del~Barrio, and G.~Botella, ``Energy-efficient mac units for fused posit arithmetic,'' in \emph{2021 IEEE 39th International Conference on Computer Design (ICCD)}, 2021, pp. 138--145.

\bibitem{2020_Guntoro_DATE}
A.~Guntoro, C.~De~La~Parra, F.~Merchant, F.~De~Dinechin, J.~L. Gustafson, M.~Langhammer, R.~Leupers, and S.~Nambiar, ``Next generation arithmetic for edge computing,'' in \emph{2020 Design, Automation \& Test in Europe Conference \& Exhibition (DATE)}, 2020, pp. 1357--1365.

\bibitem{2017_Stillmaker_scaling}
\BIBentryALTinterwordspacing
A.~Stillmaker and B.~Baas, ``Scaling equations for the accurate prediction of cmos device performance from 180nm to 7nm,'' \emph{Integration}, vol.~58, pp. 74--81, 2017. [Online]. Available: \url{https://www.sciencedirect.com/science/article/pii/S0167926017300755}
\BIBentrySTDinterwordspacing

\bibitem{2018_Carmichael_ArXiv}
Z.~Carmichael, H.~F. Langroudi, C.~Khazanov, J.~Lillie, J.~L. Gustafson, and D.~Kudithipudi, ``Deep positron: A deep neural network using the posit number system,'' in \emph{2019 Design, Automation \& Test in Europe Conference \& Exhibition (DATE)}, 2019, pp. 1421--1426.

\bibitem{2019_Mach_VLSI-SOC}
S.~Mach, F.~Schuiki, F.~Zaruba, and L.~Benini, ``A 0.80pj/flop, 1.24tflop/sw 8-to-64 bit transprecision floating-point unit for a 64 bit risc-v processor in 22nm fd-soi,'' in \emph{2019 IFIP/IEEE 27th International Conference on Very Large Scale Integration (VLSI-SoC)}, 2019, pp. 95--98.

\bibitem{2024_Bertaccini_IEEE}
L.~Bertaccini, G.~Paulin, M.~Cavalcante, T.~Fischer, S.~Mach, and L.~Benini, ``Minifloats on risc-v cores: Isa extensions with mixed-precision short dot products,'' \emph{IEEE Transactions on Emerging Topics in Computing}, vol.~12, no.~4, pp. 1040--1055, 2024.

\end{thebibliography}

\end{document}